\documentclass[11pt]{article}

\usepackage{amssymb,amsmath,amsthm}
\usepackage[totalwidth=17cm,totalheight=24cm]{geometry}
\usepackage{graphicx}

\newcommand{\C}{{\mathbb C}}
\newcommand{\R}{{\mathbb R}}
\newcommand{\Z}{{\mathbb Z}}
\newcommand{\Oc}{{\mathbb O}}
\newcommand{\Hq}{{\mathbb H}}

\newcommand{\im}{{\rm i }}

\newcommand{\be}{\begin{eqnarray}}
\newcommand{\ee}{\end{eqnarray}}
\newcommand{\id}{{\mathbb I}}

 \theoremstyle{plain}% default
  \newtheorem{theorem}{Theorem}

  \theoremstyle{definition}

  \theoremstyle{remark}

\begin{document}

\title{${\rm SO}(9)$ characterisation of the Standard Model gauge group}
\author{Kirill Krasnov\\ {}\\
{\it School of Mathematical Sciences, University of Nottingham, NG7 2RD, UK}}

\date{v3: November 2020}
\maketitle

\begin{abstract} 

A recent series of works by M.~Dubois-Violette, I.~Todorov and S.~Drenska characterised the SM gauge group $G_{\rm SM}$ as the subgroup of ${\rm SO}(9)$ that, in the octonionic model of the later, preserves the split $\Oc=\C\oplus\C^3$ of the space of octonions into a copy of the complex plane plus the rest. This description, however, proceeded via the exceptional Jordan algebras $J_3(\Oc), J_2(\Oc)$ and and this sense remained indirect. One of the goals of this paper is to provide as explicit description as possible and also clarify the underlying geometry. The other goal is to emphasise the role played by different complex structures in the spaces $\Oc$ and $\Oc^2$. We provide a new characterisation of $G_{\rm SM}$: The group $G_{\rm SM}$ is the subgroup of ${\rm Spin}(9)$ that commutes with of a certain complex structure $J_R$ in the space $\Oc^2$ of ${\rm Spin}(9)$ spinors. The complex structure $J_R$ is parametrised by a choice of a unit imaginary octonion. This characterisation of $G_{\rm SM}$ is essentially octonionic in the sense that $J_R$ is restrictive because octonions are non-associative. The quaternionic analog of $J_R$ is the complex structure in the space $\Hq^2$ of ${\rm Spin}(5)$ spinors that commutes with all ${\rm Spin}(5)$ transformations. 
\end{abstract}

\section{Introduction}

A recent paper \cite{Dubois-Violette:2018wgs}, continuing the works \cite{Dubois-Violette:2016kzx}, \cite{Todorov:2018mwd}, provided a new characterisation of the Standard Model (SM) gauge group, as the subgroup of transformations from ${\rm SO}(9)$ that, in the octonionic model of the later, preserve the decomposition $\Oc=\C\oplus\C^3$. The description in these papers, however, proceeded via the Jordan algebras with octonionic entries, and in this sense remained indirect. In particular, the statement that the SM gauge group appears precisely in this way relied on results from \cite{Yokota}, and  remained somewhat of a mystery as no direct verification was provided. Also, these papers do not discuss the fact that a decomposition $\Oc=\C\oplus\C^3$ is equivalent to a choice of a unit imaginary octonion $i\in{\rm Im}\,\Oc, |i|^2=1$. This fact, together with the observation in \cite{Dubois-Violette:2018wgs}, gives a strikingly simple characterisation of the SM gauge group as a subgroup of ${\rm SO}(9)$ that preserves the structures induced on $\Oc$ by a choice of a unit imaginary octonion $i\in{\rm Im}\,\Oc$. 

Further, it is well-known that a unit imaginary octonion $i\in{\rm Im}$ gives rise to a complex structure on $\Oc$. What is less often emphasised, however, is that there are two natural complex structures arising, those given by left and right multiplication by $i$. Further, when one wants to extend this to the space of ${\rm Spin}(9)$ spinors, which is $\Oc^2$, there are even more possibilities. Each of the arising complex structures on $\Oc^2$ then has its own centraliser. Our main new result is that there is exists a complex structure whose centraliser is precisely the SM gauge group. This ${\rm Spin}(9)$ complex structure has a quaternionic ${\rm Spin}(5)$ analog. In the quaternionic case this is the complex structure that commutes with all ${\rm Spin}(5)$ transformations. 

It is the simplicity of this statement that prompted us to write this paper. It is standard in the context of Grand Unified Theories (GUT) to put the SM gauge group $G_{\rm SM}$ inside a bigger, and in particular simple Lie group, with the suitable GUT groups $G_{\rm GUT}$ being ${\rm SU}(5), {\rm SO}(10)$ and $E_6$. However, there is never an explanation of what makes $G_{\rm SM}$ special inside the GUT group. One simply adds appropriate Higgs fields to break the GUT symmetry down to $G_{\rm SM}$, and then further to break the electroweak symmetry. But there is never a hint of an explanation of what makes $G_{\rm SM}$ special. While the group ${\rm SO}(9)$ is clearly unsuitable as a GUT group because its spinor representation $\Oc^2=\R^{16}=\C^8$ can only describe (the left-handed, see below) half of the SM fermions of one generation, it is intriguing that the SM gauge group does fit inside ${\rm SO}(9)$. Moreover, we find it striking that the breaking down from ${\rm SO}(9)$ to $G_{\rm SM}$ is caused just by making a choice of a complex structure on ${\rm SO}(9)$ spinors. In turn, this choice is parametrised by a single unit imaginary octonion. In particular, this explains how a complex gauge group $G_{\rm SM}$ can arise from the real ${\rm SO}(9)$. 

The observation of \cite{Dubois-Violette:2018wgs}, as well as our observation emphasising the role played by a choice of complex structure in $\Oc^2$, does not constitute a new GUT model.\footnote{We note, however, that the paper \cite{Dubois-Violette:2018wgs} does make further steps in the direction of constructing a field theory model.} But these observations are tantalising, and suggest that there maybe similar mechanisms at play in the more realistic ${\rm SO}(10)$ GUT setting. The ${\rm SO}(10)$ setup will be revisited elsewhere. 

For the convenience of the reader we describe the main elements of our analysis already here in the Introduction. The group ${\rm SO}(9)$ has the group ${\rm SO}(8)$ as a subgroup. The triality automorphism of ${\rm SO}(8)$ implies that this group can be described in octonionic terms. In fact, the triality for ${\rm SO}(8)$ {\it explains} why octonions exist. Indeed, the triality permutes the three different 8-dimensional representations of ${\rm Spin}(8)$. Further, for any orthogonal group (in even dimensions) we have the map
\be
V\times S_+ \to S_-,
\ee
where $V$ is the vector representation and $S_\pm$ are the chiral (or Weyl) spinor representations. This map is given by the Clifford multiplication. For ${\rm Spin}(8)$, the triality map identifies $V,S_\pm$ are thus gives a map 
\be
V\times V\to V.
\ee
This map identifies $V$ with $\Oc$ with its octonionic multiplication. This construction also implies that ${\rm Spin}(8)$ can be given on octonionic description. This description extends to ${\rm Spin}(9)$, and is as follows. 

The Clifford algebra ${\rm Cl}_9$ is described as the algebra of matrices in ${\rm End}(\Oc\oplus\Oc)$ of the type 
\be\label{type-intr}
X(r, {\bf x}) = \left( \begin{array}{cc} r & L_{\bf x} \\ L_{\overline{{\bf x}}} & - r \end{array}\right), \qquad r\in \R, {\bf x}\in \Oc.
\ee
Here $L_{\bf x}$ is the operator of {\it left} multiplication by ${\bf x}\in \Oc$. An easy calculation then shows that $X(r, {\bf x})^2=(r^2+|{\bf x}|^2)\id$, and so these matrices generate ${\rm Cl}_9$. The group ${\rm Spin}(9)$ can then be described as generated by an even number of reflections in $\R^9$. In turn, a reflection on a unit vector is again described in terms of matrices of the type (\ref{type-intr}). A reflection of $(r,{\bf x})\in(\R,\Oc)=\R^9$ on a unit vector $(s,{\bf y})\in \R^9$ is described as
\be\label{reflection-intr}
X(r, {\bf x}) \to - X(s, {\bf y}) X(r, {\bf x}) X(s, {\bf y}), \qquad s^2 + |{\bf y}|^2=1.
\ee
This reflection is defined as the transformation that changes the sign of the component of $(r,{\bf x})$ along $(s,{\bf y})$, and leaves the orthogonal complement component unchanged. This gives a complete characterisation of ${\rm Spin}(9)$.

Let us now consider structures arising by choosing a unit imaginary octonion 
\be
i\in {\rm Im}\,\Oc, \, |i|^2=1.
\ee
This equips $\Oc$ with {\it two} (commuting) complex structures 
\be
J_L := L_i, \qquad J_R: = R_i,
\ee
where $L_i (R_i)$ is the left (right) multiplication by $i$. Indeed, $J_L^2=L_i L_i = L_{i^2}=-\id$, and similarly for $J_R$. It is also clear that these two complex structures commute. Their product 
\be\label{I}
I = J_L J_R 
\ee
is a linear map on $\Oc$ that squares to plus the identity $I^2=\id$. As such, it has eigenvalues $\pm 1$. It is easy to check that the subspace ${\rm Span}(\id, i)\subset \Oc$ is the eigenspace of eigenvalue $-1$ for $I$, while the remaining 6-dimensional subspace ${\rm Span}(\id, i)^\perp\subset \Oc$ is the eigenvalue $+1$ eigenspace. In other words, $I$ is the reflection on the $\id, i$ plane in $\Oc$. 

Thus, a choice of a unit imaginary octonion allows to describe $\Oc$ as $\C^4$. Moreover, this is possible in two different ways, by using $J_{L,R}$. The difference between these two complex structures is measured by the operator (\ref{I}). Each of these two complex structures identifies $\Oc\sim \C^4$, but this occurs in a different way. Only a subset $\C\subset \C^4$ is considered to be $(1,0)$ with respect to both of these complex structures, while $J_{L,R}$ disagree as to whether $\C^3\subset \C^4$ is $(1,0)$ or $(0,1)$. Thus, both of these complex structures together single out a preferred copy of the complex plane and provide an identification
\be\label{structure}
\Oc=\C\oplus\C^3.
\ee
Thus, the structure of the split (\ref{structure}) that plays the essential role in the characterisation \cite{Dubois-Violette:2018wgs} is equivalent to a choice of two commuting complex structures $J_{L,R}$ in $\Oc$, or, equivalently, one of these complex structures, and a commuting with it para-complex structure $I: I^2=\id$. 

We now remark, to be useful later, that the subgroup of ${\rm Spin}(8)$ that commutes with $J_L$ is ${\rm SU}(4)$. As follows from the analysis of Section 3, the subgroup of ${\rm Spin}(8)$ that commutes with the action of $J_R$ on spinors of both chiralities is $S({\rm U}(1)\times{\rm U}(3))$, which is one of the factors that appears in the Standard Model gauge group
\be\label{GSM}
G_{\rm SM}={\rm SU}(3)\times{\rm SU}(2)\times{\rm U}(1)/\Z_6.
\ee

We now consider the similar setup for ${\rm Spin}(9)$. This acts on 2-component columns with entries in $\Oc$. We extend the complex structures $J_{L,R}$ from $\Oc$ to $\Oc^2$ via
\be\label{J-LR}
J_L = \left( \begin{array}{cc} L_i & 0 \\ 0 & L_i \end{array}\right), \qquad J_R = \left( \begin{array}{cc} R_i & 0 \\ 0 & R_i \end{array}\right).
\ee
In fact, there are choices that are made here, as one could introduce a minus sign when acting on the second copy of $\Oc$. One can also consider mixed complex structures that use left multiplication for one of the copies of $\Oc$, and the right multiplication for the other. We will discuss these choices below, but it is important to keep in mind that each such choice has its distinct centraliser in ${\rm SO}(9)$.  It is the two choices indicated that are particularly relevant for the purpose of seeing the SM gauge group emerging. The SM gauge group arises as the centraliser of $J_R$, but it is also instructive to consider the case of $J_L$. 

The transformations from ${\rm Spin}(9)$ that commute with $J_L$ are of two types. First, there are transformations generated by matrices of the type (\ref{type-intr}) with ${\bf x}$ commuting with $i$. This forces ${\bf x}$ to lie in the copy of the complex plane generated by $1,i$, and matrices of the type (\ref{type-intr}) with ${\bf x}\in\C$ generate ${\rm SU}(2)\sim{\rm SO}(3)$. This ${\rm SO}(3)$ sits diagonally in ${\rm SO}(9)$, as the group of rotations of 3 of the 9 directions. The subgroup of ${\rm SO}(9)$ that commutes with this ${\rm SO}(3)$ is ${\rm SO}(6)\sim{\rm SU}(4)$. Overall, the subgroup of ${\rm Spin}(9)$ that commutes with $J_L$ is ${\rm SU}(2)\times {\rm SU}(4)/\Z_2$. 

The transformations from the described in the previous paragraph ${\rm SU}(2)$ also commute with $J_R$. However, not all of the transformations from the ${\rm SU}(4)$ commute with $J_R$. Analysis described in Section 3 shows that the relevant transformations are from $S({\rm U}(1)\times{\rm U}(3))\subset{\rm SU}(4)$. Putting these elements together, we 
get the main statement of this paper
\begin{theorem} The subgroup of transformations in ${\rm Spin}(9)$ that commutes with the complex structure $J_R$ given in (\ref{J-LR}) is the Standard Model gauge group (\ref{GSM}). 
\end{theorem}
What is most striking about this result is that we need only one complex structure on $\Oc^2$ to characterise $G_{\rm SM}$, not two. Indeed, the characterisation \cite{Dubois-Violette:2018wgs} describes $G_{\rm SM}$ as the subgroup of ${\rm Spin}(9)$ that preserves the structure (\ref{structure}) on $\Oc$. We have discussed that such a structure is equivalent to a choice of two commuting complex structures on $\Oc$. Thus, Theorem 1 gives a stronger characterisation of $G_{\rm SM}$ than the one in \cite{Dubois-Violette:2018wgs} in the sense that less structure is needed to see $G_{\rm SM}$ emerging.

\noindent{\bf Comparison with previous work.} The papers \cite{Dubois-Violette:2016kzx} and \cite{Todorov:2018mwd} observed that $G_{\rm SM}$ can be characterised as the intersection of two maximal subgroups ${\rm SO}(9)$ and ${\rm SU}(3)\times{\rm SU}(3)/\Z_3$ of the exceptional Lie group $F_4$. Moreover, the group $F_4$ is intimately related to the octonions (as all exceptional Lie groups are, see e.g. \cite{Baez:2001dm}). In particular, the Lie algebra of $F_4$ can be realised as the Lie algebra of ${\rm SO}(9)$ plus its spinor representation
\be
{\mathfrak f}_4 = {\mathfrak so}(9) + S.
\ee
Moreover, the 16-dimensional spinor representation $S$ of ${\rm SO}(9)$ can be naturally viewed as two copies of the octonions
\be
S=\Oc\oplus\Oc,
\ee
so that ${\rm SO}(9)$ elements become matrices with values in ${\rm End}(\Oc\oplus\Oc)$
\be\label{oct-model}
{\rm SO}(9)\subset {\rm End}(\Oc\oplus\Oc).
\ee
The final observation is that the group ${\rm SU}(3)\times{\rm SU}(3)/\Z_3$ can be characterised as the subgroup of $F_4$ that preserves a choice of a copy of the complex plane $\C$ in the octonions. These observations taken together imply that the SM gauge group can be characterised as the subgroup of ${\rm SO}(9)$ that preserves the split (\ref{structure}) when the ${\rm SO}(9)$ is realised  as acting on two copies of octonions (\ref{oct-model}). This statement appears explicitly in Section 4 of  \cite{Dubois-Violette:2018wgs}. What is new in this paper is the observation that the structure of the split (\ref{structure}) is induced just by a choice of a unit imaginary octonion. Such a choice parametrises the complex structure $J_{R}$ whose centraliser is precisely the SM gauge group. Our description also works directly with ${\rm SO}(9)$ and avoids using Jordan algebras that are so central in \cite{Dubois-Violette:2018wgs}, \cite{Dubois-Violette:2016kzx} and \cite{Todorov:2018mwd}.

There are also parallels between our story and that developed over the years by Geoffrey Dixon. This story is summarised in the book \cite{Dixon-book}. For more recent accounts see e.g. \cite{Dixon:1999rg} and \cite{Dixon:2010px}. Dixon's work is hard to read, as everything is done very explicitly and in components, which makes it very difficult to distill what the important statements are. What is clear is that Dixon emphasises the algebra $\C\times\Hq\times\Oc$, as well as the group ${\rm SO}(1,9)$. The Clifford algebra of the later is generated by $2\times 2$ matrices of the type (\ref{type-intr}) but without imposing the tracefree condition. These are the matrices that constitute the Jordan algebra $J_2^8$. The book \cite{Dixon-book} claims that the structure of the SM, to a large extent, follows from the structure of this mathematical objects, by making some choices. Thus, it is possible that some statements of this paper are also contained somewhere in Dixon, but we were unable to find them stated clearly. 

\bigskip

We now proceed to describe all elements of our construction in more details. The groups of matrices of the type (\ref{type-intr}) generates ${\rm Cliff}_3$ when ${\bf x}\in \C$, and ${\rm Cliff}_5$ when ${\bf x}\in \Hq$, the quaternions. So, there is a simpler version of the main claim for ${\rm Spin}(5)$, which we describe in  Section \ref{sec:quat}. We use the description that proceeds via the Clifford algebra, as this can be generalised to the octonions. However, in the quaternionic case a completely explicit matrix description is also possible, and for completeness we discuss it in the Appendix. The main octonionic story is explained in Section \ref{sec:oct}. We conclude with a discussion.

\section{${\rm SO}(5)$ and quaternions}
\label{sec:quat}

We describe the case of ${\rm SO}(5)$ and quaternions here. We start by describing the all-familiar case of the rotation group in three dimensions. 

\subsection{${\rm SO}(3)$ and complex numbers}

We start by explaining the baby case of our constructions. The special orthogonal group in three dimensions is famously isomorphic to ${\rm SU}(2)/\Z_2$, the group of special unitary $2\times 2$ matrices, modulo a discrete subgroup consisting of one non-trivial element. The construction that extends to octonions is the following easily recognisable characterisation of the Clifford algebra ${\rm Cl}_3$. Thus, the three-dimensional Clifford algebra is generated by the $2\times 2$ matrices
\be\label{matrix-X}
X(r, {\bf x}) = \left( \begin{array}{cc} r & L_{\bf x} \\ L_{\overline{{\bf x}}} & - r \end{array}\right), \qquad r\in \R, {\bf x}\in \C.
\ee
The notations are chosen so as to make the changes necessary when passing to quaternions and octonions minimal. Here $L_{\bf x}$ is the left multiplication, but $\C$ is commutative, and so there is no difference between the left and right multiplications. One recognises in the object $X(r,{\bf r})$ the linear combination of Pauli matrices
\be
X(r, {\bf x}) = x_1 \sigma^1 + x_2 \sigma^2 + x_3 \sigma^3,
\ee
where $r=x_3$ and ${\bf x}=x_1 -\im x_2$. The Pauli matrices $\sigma^i, i=1,2,3$ are the $\gamma$-matrices in three dimensions
\be
\sigma^i \sigma^j + \sigma^j \sigma^i = 2 \delta^{ij} \id,
\ee
and so generate the Clifford algebra ${\rm Cl}_3$. The Clifford algebra relations can also be written in terms of $X(r, {\bf x})$, and read
\be
X(r, {\bf x}) X(r, {\bf x}) = (r^2 + |{\bf x}|^2) \id. 
\ee
The group of $2\times 2$ complex matrices acts on matrices $X(r, {\bf x})$ via
\be
{\rm GL}(2,\C)\ni g: X(r, {\bf x}) \to g X(r, {\bf x}) g^\dagger,
\ee
and the subgroup that preserves the space of matrices of the form (\ref{matrix-X}) (which are Hermitian tracefree) is the subgroup ${\rm SU}(2)$ of special unitary matrices. This gives a homomorphism from ${\rm SU}(2)$ to ${\rm SO}(3)$, and this has kernel $\Z_2$ with non-trivial element $-\id\in {\rm SU}(2)$, which provides the isomorphism ${\rm SO}(3)={\rm SU}(2)/\Z_2$. 

The description of ${\rm SO}(3)$ in terms of $2\times 2$ matrices with complex entries has an extension to quaternions, but not to non-associative octonions. The description that {\it can be} generalised to octonions is as follows. The matrices of the type (\ref{matrix-X}) with $r^2+|{\bf x}|^2=1$ generate the group ${\rm Spin}(3)$ in the sense that every element of the rotation group can be represented as an even number of reflections along unit vectors in $\R^3$. The reflection along a unit vector in $\R^3$ is obtained by conjugating the matrix $X(r, {\bf x})$ by another matrix of this form, see (\ref{reflection-intr}).

Matrices of the type (\ref{matrix-X}) corresponding to a unit vector act on 2-component columns with complex entries. They generate the ${\rm Pin}(3)$ group. The group ${\rm Spin}(3)$ is the one generated by an even number of unit matrices of the type (\ref{matrix-X}), and the 2-component columns with complex entries form its spinor representation space. A very useful reference on this material on spinors and Clifford algebras (and much more) is \cite{Harvey}.

\subsection{Quaternions} 

Quaternions is a set $\Hq$ of objects that can be represented as
\be
q = x_0 + {\bf i} x_1 +{\bf j} x_2 +{\bf k} x_3, \qquad x_{0,1,2,3}\in \R,
\ee
with the units ${\bf i,j,k}$ satisfying ${\bf i}^2={\bf j}^2={\bf k}^2=-1$, and ${\bf i}{\bf j}={\bf k}=-{\bf j}{\bf i}$, as well as all the relations that follow from these. Quaternions are thus non-commutative, but associative. The conjugation is defined as
\be
\overline{q}:=  x_0 - {\bf i} x_1 -{\bf j} x_2 -{\bf k} x_3,
\ee
the operation that changes the sign of each of the three imaginary unit. We then have
\be
q\overline{q} = (x_0)^2+(x_1)^2+(x_2)^2+(x_3)^2 \equiv |q|^2.
\ee
The right-hand-side of the product $q\overline{q}$ is real in the sense that it does not involve any of the imaginary units. This allows to define the operation of division by a quaternion $q^{-1} = \overline{q}/|q|^2$, and so $\Hq$ is a division algebra. It is also a composition algebra with the norm satisfying $|p q|^2=|p|^2 |q|^2$. 

\subsection{Complex structure(s) on $\Hq$}

Quaternions acts on themselves by the operation of, say, left multiplication. It is then easy to see that choosing an imaginary unit quaternion $i\in{\rm Im}\,\Hq, |i|^2=1$ equips $\Hq$ with a complex structure. Indeed, let $L_i$ be the operation of left multiplication by $i$. Then for a unit imaginary quaternion $\overline{i}=-i$ and we have 
\be
L_i^2 = - L_i L_{\overline{i}} = - L_{i\overline{i}} = -\id.
\ee
We thus see that different complex structures on $\Hq$ are parametrised by points on the two-sphere $S^2\subset \R^3={\rm Im}\,\Hq$ in the space of imaginary quaternions. 

For example, let us take $i={\bf i}$. Such a choice splits $\Hq$ into two copies of complex plane $\Hq=\C\oplus\C$. The $(1,0)$ subspace of $\Hq_\C$ is the one on which $J_L= L_{\bf i}$ has eigenvalue $-\im$. It is spanned by the vectors
\be\label{ab}
1+\im {\bf i}, \qquad {\bf j}+\im {\bf k}. 
\ee

We cold instead take the complex structure given by the right multiplication $J_R=R_{\bf i}$ by ${\bf i}$. This complex structure agrees about $1+\im {\bf i}$ being a $(1,0)$ vector, but disagrees with $L_{\bf i}$ about ${\bf j}+\im {\bf k}$. With respect to $R_{\bf i}$ this is a $(0,1)$ vector. 

Thus $J_L$ identifies $\Hq=\C^2$, while both $J_{L,R}$ select a preferred choice of the complex plane in $\C^2$ and provide the split
\be
\Hq=\C\oplus\C.
\ee

\subsection{${\rm SO}(5)$ Lie algebra}

There is a matrix description of ${\rm SO}(5)$, see Appendix. This description, however, does not extend to the case of octonions. But its Lie algebra version does extend, and so we describe it now to help understand the octonionic case. 

The Clifford algebra ${\rm Cl}_5$ is generated by the matrices of the type (\ref{matrix-X}), but with ${\bf x}\in\Hq$. Explicitly, if we describe $\Hq=\R^4$, we get the following $8\times 8$ real matrices 
\be
\gamma^i = \left( \begin{array}{cc} 0 & E^i \\ -E^i  & 0 \end{array}\right), \quad  \gamma^4= \left( \begin{array}{cc} 0 & \id \\ \id  & 0 \end{array}\right), \quad \gamma^5 = \left( \begin{array}{cc} \id & 0 \\ 0 & -\id \end{array}\right),
 \ee
 where the matrices $E^i$ are $4\times 4$ and are worked out as follows
 \be
 {\bf i}(x_0 + {\bf i} x_1 +{\bf j} x_2 +{\bf k} x_3) = {\bf i}x_0 -  x_1 +{\bf k} x_2 - {\bf j} x_3 \Rightarrow E^1 \left( \begin{array}{c} x_0 \\ x_1 \\ x_2 \\ x_3\end{array}\right) = \left( \begin{array}{c} -x_1 \\ x_0 \\ -x_3 \\ x_2\end{array}\right) ,
 \ee
 and so
 \be
 E^1 = -E_{12}-E_{34},
 \ee
 where $E_{ij}$ is a $4\times 4$ anti-symmetric matrix with $+1$ on the $i$th row and $j$th column. Matrices $E_{ij}$ are the generators of ${\mathfrak so}(4)$. Similarly, we have
 \be
 E^2 = -E_{13}+E_{24} ,\qquad E^3 = -E_{14}-E_{23}.
 \ee
 The matrices introduced satisfy the correct quaternionic relations
 \be
 (E^i)^2=-\id, \qquad E^1 E^2 = E^3 = - E^2 E^1.
 \ee

The Lie algebra of ${\rm SO}(5)$ is generated by the commutators of the above $\gamma$-matrices.  These are easily worked out, and the general Lie algebra element is of the form
\be\label{so5}
\left( \begin{array}{cc} A & L_{\bf x} \\ -L_{\bar{\bf x}} & A' \end{array}\right),
\ee
where $A$ is an ${\mathfrak so}(4)$ matrix and $A'$ is its representation on the Weyl spinor of opposite chirality
\be
A= (a_i +b_i) E^i, \qquad A'=(a_i - b_i) E^i,\qquad a_i,b_i\in \R,
\ee
and 
\be
L_{\bf x} = x_0 \id + x_i E^i, \qquad -L_{\bar{\bf x}} = - x_0  \id + x_i E^i, \qquad x_0\in \R, x_i\in \R.
\ee
This is an explicit parametrisation of the Lie algebra ${\mathfrak so}(5)$ by three $\R^3$ vectors $a_i,b_i, x_i$ and one real number $x_0$. 

\subsection{Complex structures on $\Hq^2$ and their centralisers}

We have seen that there are two different complex structures $J_{L,R}$ on $\Hq$ arising from a choice of a unit imaginary quaternion ${\bf i}$. These can be extended to $\Hq^2$ in many different ways, and it is instructive to work out the centraliser for each choice. 

Let us start with the choice that is going to be of most relevance below, namely 
\be\label{JL-so5}
J_L = \left( \begin{array}{cc} L_{\bf i} & 0 \\ 0 & L_{\bf i} \end{array}\right). 
\ee
The requirement that it commutes with (\ref{so5}) reduces to the requirement that ${\bf i}$ commutes with ${\bf x}$, which means that ${\bf x}\in {\rm Span}(1,{\bf i})$, and that both $A,A'$ commute with $L_{\bf i}=E^1$. This means that $A=(a_1+b_1)E^1, A'=(a_1-b_1)E^1$. The centraliser of $J_L$ is thus four-dimensional, and is given by ${\mathfrak so}(3)\oplus{\mathfrak so}(2)$, where the first factor describes rotations in the $5,4,{\bf i}$ plane, and the second factor rotates the remaining two directions ${\bf j},{\bf k}$. 

Let us now describe a different complex structure. Define
\be\label{JL-prime-so5}
J_L' = \left( \begin{array}{cc} L_{\bf i} & 0 \\ 0 & -L_{\bf i} \end{array}\right).
\ee
Thus, we introduce a relative minus sign between the two copies of $\Hq$. The requirement that this commutes with (\ref{so5}) now means that $L_{\bf i}$ commutes with both $A,A'$, which again implies $A=(a_1+b_1)E^1, A'=(a_1-b_1)E^1$. However, for the off-diagonal elements the implication is now that ${\bf i}$ anti-commutes with ${\bf x}$, which means ${\bf x}\in{\rm Span}({\bf j},{\bf k})$. Again, we see that the centraliser is ${\mathfrak so}(3)\oplus{\mathfrak so}(2)$, but this is now a very different subalgebra. Inspection shows that the first factor is now the rotation in the $5,{\bf j},{\bf k}$ plane, while the remainder is the rotation in the $1,{\bf i}$ plane. 

Let us now discuss the complex structure that is built from the right multiplication. We define
\be\label{JR-so5}
J_R = \left( \begin{array}{cc} R_{\bf i} & 0 \\ 0 & R_{\bf i} \end{array}\right). 
\ee
The requirement that a Lie algebra element commutes with $J_R$ now implies that $A,A'$ commute with $R_{\bf i}$ and that $L_{\bf x}$ commutes with $R_{\bf i}$. However, the latter is true for any ${\bf x}$, and so gives no conditions. For the former, this is also true for any ${\mathfrak so}(4)$ Lie algebra element, because $E^{1,2,3}$ are given by the left multiplication with ${\bf i},{\bf j},{\bf k}$ respectively, and these commute with $R_{\bf i}$. So, all of the ${\mathfrak so}(5)$ commutes with $J_R$. This means that there exists a description of this Lie algebra in terms of complex $4\times 4$ matrices rather than real $8\times 8$ matrices. We give this description in the Appendix. The octonionic analog of $J_R$ is the complex structure whose centraliser in ${\rm Spin}(9)$ is the SM gauge group. 

We can also consider 
\be\label{JR-prime-so5}
J_R' = \left( \begin{array}{cc} R_{\bf i} & 0 \\ 0 & -R_{\bf i} \end{array}\right). 
\ee
Now the commutativity requirement reduces to the condition that $A,A'$ commute with $R_{\bf i}$, which is true for all ${\mathfrak so}(4)$. But we now need that $L_{\bf x}$ anti-commutes with $R_{\bf i}$, which is only true for ${\bf x}=0$. So, the centraliser in this case is ${\mathfrak so}(4)$. 

Let us now consider a complex structure that is built from left and right multiplication. Thus, define
\be
J_{LR} = \left( \begin{array}{cc} L_{\bf i} & 0 \\ 0 & R_{\bf i} \end{array}\right). 
\ee
Working out the commutativity implications we get that $A$ must commute with $L_{\bf i}$, which implies $A=(a_1+b_1)E^1$. There are no implications from the condition that $A'$ commutes with $R_{\bf i}$. Another non-trivial condition is $L_{\bf x} R_{\bf i} = L_{\bf i} L_{\bf x}$. This implies that ${\bf x}\in {\rm Span}(1,{\bf i})$. Overall, we see that the centraliser in this case is exactly the same as that of $J_L$. 

Finally, another remaining possibility is 
\be
J_{LR}' = \left( \begin{array}{cc} L_{\bf i} & 0 \\ 0 & -R_{\bf i} \end{array}\right). 
\ee
We again get the condition that $A$ commutes with $L_{\bf i}$, which as in the previous case implies $A=(a_1+b_1)E^1$. There is no restriction on $A'$ apart from the fact that it is related to $A$ that is already restricted. The other non-trivial condition that we get is $-L_{\bf x} R_{\bf i} = L_{\bf i} L_{\bf x}$, which implies that ${\bf x}\in{\rm Span}({\bf j},{\bf k})$. The centraliser in this case is thus the same as that of $J_L'$. Thus, the left-right complex structures do not give anything new, at least in the considered now quaternionic case. 
 
 \subsection{Subgroup preserving the split $\Hq=\C\oplus\C$}
 
We now describe the case of main interest for us. As we discussed previously, left and right multiplication by ${\bf i}$ in $\Hq$ defines the split  $\Hq=\C\oplus\C$. These two complex structures can be extended to $\Hq^2$ as (\ref{JL-so5}) and (\ref{JR-so5}). The Lie algebra of the centraliser of the first is ${\mathfrak so}(3)\oplus{\mathfrak so}(2)$. The centraliser of the second is all of ${\mathfrak so}(5)$. While only $J_L$ appears interesting in the quaternionic case, it is the complex structure $J_R$ that gives the SM gauge group in the octonionic case. 

At the group level, it is clear that the centraliser of $J_L$ in ${\rm Spin}(5)$ is ${\rm U}(1)\times{\rm SU}(2)/\Z_2$. The non-trivial centre arises because the element $(-1, -\id)\in{\rm U}(1)\times{\rm SU}(2)$ is a trivial element in ${\rm Spin}(5)$. The centraliser of $J_R$ is all of the group ${\rm Spin}(5)$.
 
\subsection{Another ${\rm U}(1)\times{\rm SU}(2)/Z_2$ in ${\rm SO}(5)$}

For completeness, we mention that there is a differently embedded ${\rm U}(1)\times{\rm SU}(2)/Z_2$ subgroup of ${\rm SO}(5)$ that should not be confused with the ones we just described. This one arises as a subgroup of ${\rm SO}(4)$ that fixes the direction $5$. As the subgroup of ${\rm SU}(2,\Hq)$, this ${\rm SO}(4)$ consists of matrices of the type
\be\label{g-diag}
g = \left( \begin{array}{cc} A & 0 \\ 0 & D \end{array}\right), \qquad A,D\in {\rm SU}(2).
\ee
It preserves the $5$ coordinate, while its action on the quaternion ${\bf x}$ is the familiar action of two copies of ${\rm SU}(2)$ on a $2\times 2$ unitary matrix
\be
{\bf x} \to A{\bf x} D^\dagger.
\ee
If we now introduce a complex structure on $\R^4$ in which $x_0+\im x_1, x_2+\im x_3$ are the (anti-)holomorphic coordinates and ask for transformations that mix these anti-holomorphic coordinates with themselves, it is clear that these are given by $D$ arbitrary special unitary, while $A$ must be diagonal
\be
A= \left( \begin{array}{cc} e^{\im\phi} & 0 \\ 0 & e^{-\im\phi} \end{array}\right).
\ee
This gives a copy of ${\rm U}(1)\times{\rm SU}(2)/Z_2$ in ${\rm SO}(4)={\rm SU}(2)\times{\rm SU}(2)/\Z_2$. It is clearly embedded very differently into ${\rm SO}(5)$, and also the spinor representation $S$ decomposes differently under this version of ${\rm U}(1)\times{\rm SU}(2)/Z_2$ as compared to what was described in the previous subsection. This version of ${\rm U}(1)\times{\rm SU}(2)/Z_2$ can be said arising from requiring the preservation of the $\Hq=\C+\C$ split but in the vector representation instead, viewed as $\R^5=\R\oplus\Hq$. 

\section{Octonionic version and the SM gauge group}
\label{sec:oct}

We now describe an analogous construction with $\Hq$ replaced with $\Oc$ everywhere. The octonions are not associative, and so there is no matrix description. Nevertheless, the Clifford algebra ${\rm Cl}_9$ is still generated by matrices of the type (\ref{matrix-X}), and this generalises what we had in the quaternionic case.  

\subsection{Octonions}

Octonions, see e.g.,  \cite{Baez:2001dm}, are objects that can be represented as linear combinations of the unit octonions $1,e^1,\ldots, e^7$
\be
x = x_0 + x_1 e^1 +\ldots + x_7 e^7, \qquad x_0, x_1, \ldots, x_7 \in \R.
\ee
The conjugation is again the operation that flips the signs of all the imaginary coefficients
\be
\overline{x} = x_0 - x_1 e^1 -\ldots - x_7 e^7,
\ee
and we have 
\be
x\overline{x}= (x_0)^2 + (x_1)^2 + \ldots + (x_7)^2 \equiv |x|^2.
\ee
Octonions $\Oc$ form a normed division algebra that satisfies the composition property $|xy|^2=|x|^2|y|^2$. The cross-products of the imaginary octonions $e^1,\ldots, e^7$ can be conveniently encoded into a 3-form in $\R^7$ that arises as 
\be
C(x,y,z)=\langle xy, z\rangle, \qquad x,y,z\in {\rm Im}\Oc,
\ee
where the inner product $\langle\cdot,\cdot\rangle$ in $\Oc$ comes by polarising the squared norm
\be
\langle x,y\rangle =  {\rm Re}(x\overline{y}), \qquad x,y\in \Oc.
\ee
One possible form of $C$ is
\be\label{C}
C = e^{567} + e^5\wedge (e^{41}-e^{23}) + e^6\wedge (e^{42}-e^{31}) + e^7\wedge (e^{43}-e^{12}),
\ee
where the notation is $e^{ijk}=e^i\wedge e^j\wedge e^k$. Octonions are non-commutative and non-associative, but alternative. The last property is equivalent to saying that any two imaginary octonions (as well as the identity) generate a subalgebra that is associative, and is a copy of the quaternion algebra $\Hq$.

\subsection{Octonionic model of ${\rm SO}(9)$}

Because ${\rm Spin}(8)\subset{\rm Spin}(9)$, and the former is "octonionic", the later can be given an octonionic description as well. This arises as the already familiar description of the Clifford algebra as that of matrices of the form (\ref{matrix-X}) with entry ${\bf x}$ in either $\C, \Hq$ or $\Oc$. When ${\bf x}\in \Oc$ we get the generators of the Clifford algebra ${\rm Cl}_9$. These act on 2-component columns with octonionic entries. The matrices (\ref{matrix-X}) corresponding to unit vectors in $\R^9$ generate the group ${\rm Pin}(9)$. The group generated by an even number of the unit matrices (\ref{matrix-X}) is the spin group ${\rm Spin}(9)$, and octonionic 2-columns form its spinor representation.

\subsection{Complex structures on $\Oc$}

As in the case of quaternions, a choice of a unit imaginary octonion gives a complex structure on $\Oc$
\be
L_i^2=-\id, \qquad i\in{\rm Im}\,\Oc, \qquad |i|^2=1,
\ee
where $L_i$ is the left multiplication by $i$. Thus, the choice of a complex structure $J=L_i$ on $\Oc$ is the choice of a point on a six-sphere $i\in S^6\subset \R^7={\rm Im}\,\Oc$. 

For example, let us choose the complex structure that corresponds to $e^4$. A simple computation then shows that the $(1,0)$ subspace of $\Oc$ is spanned by
\be
1+\im e^4, \quad e^1 +\im e^5, \quad e^2 +\im e^6, \quad e^3+\im e^7.
\ee

We can also consider the right multiplication by $e^4$ as the complex structure. The $(1,0)$ subspace is then that spanned by
\be
1+\im e^4, \quad e^1 -\im e^5, \quad e^2 -\im e^6, \quad e^3-\im e^7.
\ee

A choice of a complex structure on $\Oc$ thus allows to identify $\Oc=\C^4$. The two complex structures $J_{L,R}$ together agree on the copy of the complex plane spanned by $1,e^4$, and thus select a preferred copy of $\C$ inside $\C^4$. This produces the structure
\be\label{oct-split}
\Oc=\C\oplus\C^3.
\ee

\subsection{Lie algebra of ${\rm SO}(9)$}

As in the case of ${\mathfrak so}(5)$, the Lie algebra ${\mathfrak so}(9)$ is generated by the $\gamma$-matrices of the form
\be
\gamma^i = \left( \begin{array}{cc} 0 & e^i \\ -e^i  & 0 \end{array}\right), \quad  \gamma^8= \left( \begin{array}{cc} 0 & \id \\ \id  & 0 \end{array}\right), \quad \gamma^9 = \left( \begin{array}{cc} \id & 0 \\ 0 & -\id \end{array}\right),
 \ee
 where the $8\times 8$ matrices $e^i, i=1,\ldots, 7$ are those describing the left multiplication by unit imaginary octonions. Using the multiplication table stored by the 3-form (\ref{C}), and denoting
 \be
 L_{e^i} := E^i
 \ee
 we get 
 \begin{eqnarray}\nonumber
& E^1 &= E_{12}-E_{38}+E_{47}-E_{56} \\ \nonumber
& E^2 &= E_{13}+E_{28}-E_{46}-E_{57} \\ \nonumber
& E^3 &= E_{14}-E_{27}+E_{36}-E_{58} \\ \nonumber
& E^4 &= E_{15}+E_{26}+E_{37}+E_{48} \\ \nonumber
& E^5 &= E_{16}-E_{25}-E_{34}+E_{78} \\ \nonumber
& E^6 &= E_{17}+E_{24}-E_{35}-E_{68} \\ \nonumber
& E^7 &= E_{18}-E_{23}-E_{45}+E_{67} .
 \end{eqnarray}
 These are all $8\times 8$ real (anti-symmetric) matrices, and the octonions in the column are ordered as $1,e^1,\ldots, e^7$. 
 
 The Lie algebra of ${\rm SO}(9)$ is generated by the commutators of the $\gamma$-matrices. It is important to remark that the matrices $E^1,\ldots, E^7$ generate an associative algebra, and should not be confused with the octonions $e^i$ whose product is non-associative. The Lie algebra ${\mathfrak so}(9)$ is of the same form 
  \be\label{so9}
\left( \begin{array}{cc} A & L_{\bf x} \\ -L_{\bar{\bf x}} & A' \end{array}\right),
\ee
as in the quaternion case, with ${\bf x}\in\Oc$, and 
\be
A= \frac{1}{2}[E^i,E^j] a_{ij} + E^i b_i, \qquad A'= \frac{1}{2}[E^i,E^j] a_{ij} - E^i b_i,
\ee
where $a_{ij}, b_i$ are all real. 

\subsection{Complex structures on $\Oc^2$ and their centralisers}

We now discuss the complex structures of Section 2.5 together with their centralisers. The discussion proceeds in exact parallel with that in the case of quaternions. We use $e^4$ as the unit imaginary octonion generating the complex structures everywhere. The notations are also as in section 2.5. The main new twist to the story is that the complex structure $J_R$ that commuted with all transformations from ${\mathfrak so}(5)$ is now highly non-trivial. Its centraliser in ${\rm Spin}(9)$ is precisely the Standard Model gauge group. 

\bigskip
\noindent{\bf The case of $J_L$.} The requirement that $L_{\bf x}$ commutes with $L_{e^4}$ implies that ${\bf x}\in{\rm Span}(1,e^4)$. The requirement that $L_{\bf x}$ commutes with $A,A'$ implies that only $b_4\not=0$, and also that all $a_{4i}=0$. This means that $A,A'\in {\mathfrak so}(6)\oplus{\mathfrak so}(2)$. Altogether, the centraliser is 
\be
{\mathfrak so}(3)\oplus{\mathfrak so}(6).
\ee 
The first factor here is the Lie algebra of the group of rotations in directions $9,1,e^4$, while the other factor rotates the remaining six coordinates.

 \bigskip
\noindent{\bf The case of $J_L'$.} Now we get the requirement that $L_{\bf x}$ anti-commutes with $L_{e^4}$. This implies that ${\bf x}\in{\rm Span}(e^1,e^2,e^3,e^5,e^6,e^7)$. And as in the previous case we have $A,A'\in {\mathfrak so}(6)\oplus{\mathfrak so}(2)$. Putting all together we get the centraliser to be 
\be
{\mathfrak so}(2)\oplus{\mathfrak so}(7).
\ee The first factor here describes rotations in the $1,e^4$ plane, while the second factor describes rotations in the remaining seven coordinates. We remark that this is the only complex structure from the ones we discuss that arises as the product of two $\gamma$-matrices. Indeed, it is not hard to see that $J_L'=\gamma^4 \gamma^8$. All other complex structures we discuss cannot be represented in this way. 

\bigskip
\noindent{\bf The case of $J_R$.} In the quaternionic case this complex structure commuted with all group transformations. This is not the case now, and we shall see that its centraliser is precisely the SM Lie algebra. 

The condition we want to satisfy is that {\it both} $A,A'$ commute with $R_{e^4}$ and $L_{\bf x}$ commutes with $R_{e^4}$. Let us first analyse the latter. In the case of quaternions this was automatic, but this is not longer the case for $\Oc$ as octonions are not associative. As an operator on $\Oc$ the right multiplication with $e^4$ is represented by the following matrix
\be
R_{e^4} = E_{15}-E_{26}-E_{37}-E_{48} .
\ee
This only commutes with the identity matrix and $E^4$, due to non-associativity of the octonions. Thus, this condition forces ${\bf x}\in{\rm Span}(1,e^4)$. 

The commutant of $R_{e^4}$ of either $A$ {\it or} $A'$ in ${\mathfrak so}(8)$ is checked to be 15-dimensional. However, when one imposes the condition that {\it both} $A,A'$ commute with $R_{e^4}$, the commutant is 9-dimensional and is given by $b_i=0$ for $i\not=4$ and $a_{4i}=0$, as well as 
\be\label{a-conds}
a_{12}=a_{56}, \quad a_{23}=a_{67}, \quad a_{13}=a_{57}, \\ \nonumber
a_{27}=a_{36}, \quad a_{17}=a_{35}, \quad a_{16}=a_{25}.
\ee
This is then checked to be the Lie algebra ${\mathfrak u}(1)\oplus{\mathfrak su}(3)\subset{\mathfrak so}(6)={\mathfrak su}(4)$, where ${\mathfrak su}(3)\subset{\mathfrak so}(7)$ is the one stabilising two spinors $1,e^4$. 

The Lie subalgebra 
\be\label{so3-so9}
\left( \begin{array}{cc} E^4 b_4 & L_{\bf x} \\ -L_{\bar{\bf x}} & - E^4 b_4 \end{array}\right), \qquad {\bf x}\in{\rm Span}(1,e^4)
\ee
is that generated by matrices of the form (\ref{matrix-X}) with ${\bf x}$ in the copy of the complex plane spanned by $1,e^4$. In other words, this is the Lie algebra ${\mathfrak so}(3)$. The Lie subalgebra
\be\label{su3-so9}
\left( \begin{array}{cc} \frac{1}{2}[E^i,E^j] a_{ij} & 0 \\ 0 &  \frac{1}{2}[E^i,E^j] a_{ij} \end{array}\right), 
\ee
with $a_{4i}=0$ and the other parameters satisfying (\ref{a-conds}) is the Lie algebra of ${\mathfrak u}(1)\oplus{\mathfrak su}(3)$. The subalgebras (\ref{so3-so9}) and (\ref{su3-so9}) commute inside ${\mathfrak so}(9)$. Together, they form the centraliser of $J_R$ in ${\mathfrak so}(9)$. We thus see the Lie algebra of the SM gauge group 
\be
{\mathfrak su}(2)\oplus{\mathfrak u}(1)\oplus{\mathfrak su}(3)
\ee
arising as the centraliser of a single complex structure $J_R$. 

\bigskip
\noindent{\bf The case of $J_R'$.} This case is also interesting. The conditions that $A,A'$ commute with $R_{e^4}$ again give the transformations from (\ref{su3-so9}) together with 
\be
\left( \begin{array}{cc} E^4 b_4 & 0 \\ 0 & - E^4 b_4 \end{array}\right).
\ee
The remaining condition is that $-L_{\bf x} R_{e^4} = R_{e^4} L_{\bf x}$. This is never satisfied, so ${\bf x}=0$. The centraliser is then ${\mathfrak so}(2)$ generating rotations on the $1,e^4$ plane, as well as the subalgebra ${\mathfrak u}(1)\oplus{\mathfrak su}(3)$ of ${\mathfrak so}(6)={\mathfrak su}(4)$, the latter describing rotations in the $e^{1,2,3,5,6,7}$ subspace. Overall, we get
\be
{\mathfrak u}(1)\oplus{\mathfrak u}(1)\oplus{\mathfrak su}(3)
\ee
as the centraliser in this case. Thus, only the diagonal ${\mathfrak u}(1)$ of the weak algebra ${\mathfrak su}(2)$ remains unbroken in this case. 

\bigskip
\noindent{\bf The cases of $J_{LR}, J_{LR}'$.}  Let us now consider the complex structure that uses both left and right multiplication operators. The arising conditions are now that $A$ commutes with $L_{e^4}$, and $A'$ commutes with $R_{e^4}$. The first of these conditions reduces to $A\in{\mathfrak so}(6)$. The second condition the further reduces $A\in {\mathfrak u}(1)\oplus{\mathfrak su}(3)$. The other conditions that must be satisfied are $L_{\bf x} R_{e^4}=L_{e^4} L_{\bf x}$ and a similar condition for $L_{\bar{\bf x}}$. This is never satisfied, so ${\bf x}=0$ and the centraliser in this case is
\be
{\mathfrak u}(1)\oplus{\mathfrak su}(3).
\ee
Thus, the weak group is fully broken in this case. 

The case of $J_{LR}'$ is similar. The conditions on ${\bf x}$ can only be satisfied for ${\bf x}=0$, while the conditions on the $A,A'$ remain unchanged. So, the centraliser of $J_{LR}'$ is the same as that of $J_{LR}$. 

\subsection{The embedding of the SM gauge group}

We have seen the structure of the SM gauge group arising at the level of the Lie algebra, as the commutant of $J_R$. Let us now discuss what the corresponding gauge group is. We need to determine the transformations from ${\rm U}(1)\times{\rm SU}(3)\times{\rm SU}(2)$ that correspond to trivial transformations inside ${\rm SO}(9)$. We embed ${\rm U}(1)\times{\rm SU}(3)$ into ${\rm SU}(4)\sim{\rm SO}(6)$ as
\be
{\rm U}(1)\times{\rm SU}(3)\ni  (\alpha,g) \to \left( \begin{array}{cc} \alpha^{-3} & 0 \\ 0 & \alpha g \end{array}\right) \in {\rm SU}(4).
\ee
The spinor representation $\Oc^2=\R^{16}=\C^8$ of ${\rm SO}(9)$ splits into two complex representations, plus their complex conjugates. Under the above embedding of ${\rm U}(1)\times{\rm SU}(3)$ and ${\rm SU}(2)$ these transform as follows
\be
L= ({\bf 1}, {\bf 2})_{-1}, \qquad Q=({\bf 3}, {\bf 2})_{1/3}.
\ee
where the numbers in brackets indicate the dimensions of the ${\rm SU}(3)$ and ${\rm SU}(2)$ representations respectively. The subscript is the ${\rm U}(1)$ charge. In other words, we get two ${\rm SU}(2)$ doublets only one of which transforms under ${\rm SU}(3)$. The explicit transformation law is
\be
{\rm U}(1)\times{\rm SU}(3)\times {\rm SU}(2) \ni (\alpha,g,h) : L\to \alpha^{-3} h L, \qquad Q\to \alpha g h Q.
\ee
These are indeed the correct transformation laws of the left-handed fermions in the SM, with this ${\rm U}(1)$ being the hypercharge. The right-handed fermions that are ${\rm SU}(2)$ singlets are not part of $\Oc^2$ on which ${\rm SO}(9)$ acts. So, the spinor of ${\rm SO}(9)$ can at best model the left-handed particles. This is of course as expected because there is simply no room for the right-handed particles. Together with the left-handed ones they need a copy of $\C^{16}$. What we have here is a copy of $\Oc^2=\R^{16}=\C^8$, and not $\C^{16}$. But the SM gauge group is adequately described. 

One can then easily see that the element
\be
( e^{\pi \im/3}, e^{2\pi \im/3} \id, -\id) \in {\rm U}(1)\times{\rm SU}(3)\times {\rm SU}(2)
\ee
generates a normal subgroup $\Z_6$ whose elements do not act on $L,Q$. Thus, the true gauge group that results from this construction is indeed
\be
G_{\rm SM}={\rm U}(1)\times{\rm SU}(3)\times {\rm SU}(2) /\Z_6.
\ee

\subsection{Automorphism $\omega$ of order three}

This subsection arose as a result of communication with the authors of \cite{Todorov:2018mwd}. As is explained in book \cite{Yokota}, see Section 2.12, a choice of an imaginary octonion, apart from providing the split $\Oc=\C\oplus\C^3$, also selects a subgroup $\Z_3\subset{\rm Aut}\,\Oc=G_2$. The SM gauge group can then be described as the subgroup of transformations in ${\rm Spin}(9)$ that commute with the generator $\omega$ of this $\Z_3$
\be
G_{\rm SM} = ({\rm Spin}(9))^\omega.
\ee

In more details, choosing an imaginary unit octonion ($e^4$ in the description above) allows to write $\Oc=\Hq_\C$ so that any octonion can be written as
\be
x = z + Z^k e_k, \qquad z, Z^k \in \C, \quad k=1,2,3,
\ee
and the generators $e_k$ satisfy the quaternion algebra $e_k e_l = -\delta_{kl} + \epsilon_{klm} e_m$ and anti-commute with the imaginary unit $e^4\equiv i$. We then have the following element
\be
\omega = - \frac{1}{2} + \frac{\sqrt{3}}{2} i \in \C\subset \Oc.
\ee
Its action on $\Oc$ is given by
\be
\omega(z+ Z^k e_k) = z + \omega Z^k e_k,
\ee
which is the diagonal action of $\omega\in\C$ on $(Z^1,Z^2,Z^3)\in \C^3$. One has $\omega^3=\id$. The claim is then that the subgroup $(G_2)^\omega$ of the group of automorphisms of the octonions that commute with $\omega$ is ${\rm SU}(3)$, into which $\Z_3$ generated by $\omega$ is embedded as its centre. Another claim is that the subgroup $({\rm Spin}(9))^\omega$ of transformations in ${\rm Spin}(9)$ that commute with $\omega$ is the Standard Model gauge group. This gives an alternative way of saying that the SM gauge group is a subgroup of ${\rm Spin}(9)$ that results from singling out an imaginary unit octonion. We believe our description in terms of a single complex structure $J_R$ is simpler. 

\subsection{Split signature version}

For completeness, we comment on the split signature case, which is completely analogous. Replacing $\Hq$ with the split quaternions $\Hq'$ we generate the Clifford algebra ${\rm Cliff}_{3,2}$, and then the group ${\rm Spin}(3,2)$. The subgroup of this that commutes with $J_L$ is still ${\rm SU}(2)\times{\rm U}(1)/\Z_2$. The first factor rotates the first three directions, while the second rotates the remaining two. 

Working with the split octonions we generate the group ${\rm Spin}(5,4)$. The subgroup of this that commutes with $J_L$ is ${\rm SU}(2)\times{\rm SU}(2,2)/\Z_2$. The subgroup that commutes with $J_R$ is the non-compact analog of the SM gauge group ${\rm SU}(2)\times {\rm U}(1)\times {\rm SU}(1,2)/\Z_6$. 

\section{Dicussion}

The most important lesson from the construction described, as well as those in works \cite{Dubois-Violette:2018wgs}, \cite{Dubois-Violette:2016kzx}, \cite{Todorov:2018mwd}, as well as \cite{Todorov:2018yvi} and \cite{Todorov:2019hlc}, is that Standard Model seems to know about the octonions. This is of course a statement with a history, see \cite{Gunaydin:1973rs} as well as works by G. Dixon, but the new twist described here is that there is now a simple and elegant characterisation of the SM gauge group as a subgroup of ${\rm Spin}(9)$ that commutes with a certain complex structure on the space of its spinors. The complex structure $J_R$ in question is restrictive only for octonions, in the sense that its quaternion analog commutes with all of ${\rm Spin}(5)$. Only for octonions, because of their non-associativity, the centraliser of $J_R$ is a subgroup of ${\rm Spin}(9)$. This makes it clear that the described mechanism that breaks ${\rm Spin}(9)$ to $G_{\rm SM}$ is octonionic in nature. In the absence of any better understanding of the patterns visible in the SM, any construction that singles out the SM gauge group should be taken seriously. 

There are two natural questions that arise in relation to our construction. The first one is whether the provided characterisation of the SM gauge group as a subgroup of ${\rm SO}(9)$ can be used to write a new field theory model. It is clear that ${\rm SO}(9)$ is not sufficient for such a purpose, because the space of ${\rm Spin}(9)$ spinors $\Oc^2=\C^8$ is too small. One has to consider ${\rm Spin}(10)$ instead, whose Weyl spinor is $\C^{16}$ and is sufficient to describe all spinors of one SM generation. Because ${\rm Spin}(8)\subset{\rm Spin}(10)$, there also exists an octonionic description of the latter. This description is explained in \cite{Bryant}. It is a natural question whether there is also a characterisation of the SM gauge group inside ${\rm Spin}(10)$ along the same lines as was described here. We will address this question elsewhere. 

The other natural question is mathematical, and is whether there are Riemannian manifolds whose holonomy group exhibits reduction along the pattern described in this article. Thus, it is known for example that there are 16-dimensional manifolds whose holonomy group ${\rm SO}(16)$ is reduced to ${\rm SO}(9)$. These are the versions of the octonionic projective plane $F_4/{\rm SO}(9)$. One can then ask whether there are examples of manifolds, perhaps also in 16 dimensions, where the holonomy is reduced further and becomes that valued in the SM gauge group. This is along the lines of the characterisation of the SM gauge group explained in \cite{Baez:2005yf}. This reference pointed out that the SM gauge group, as a subgroup of ${\rm SU}(5)\subset{\rm SO}(10)$, is precisely the holonomy group of the product of two Calabi-Yau manifolds of complex dimensions $2,3$. The fact that the SM gauge group is naturally a subgroup of ${\rm SO}(9)$ and this has a natural action in the tangent space of a 16-dimensional manifold (via its spinor representation) suggests that there may also be a similar geometric description of $G_{\rm SM}$. It would be interesting to find it. 

\section*{Appendix: ${\rm SO}(5)$ as ${\rm SU}(2,\Hq)$}

In the case of quaternions a more explicit matrix description of the above story as available. We describe it for completeness. 

\subsection{Quaternions as unitary matrices}

We choose a unit imaginary quaternion and use $J_R$ to identify $\Hq=\C^2$. Then the left multiplication $L_p$ by $p\in \Hq$ commutes with $J_R$. It can therefore be described in matrix terms, as a $2\times 2$ complex matrix acting on $\C^2$. Working out the details shows that a quaternion can be encoded into a $2\times 2$ unitary matrix
\be\label{quat-matrix}
q = \left( \begin{array}{cc} a & b \\ - b^* & a^* \end{array}\right), 
\ee
where $a,b$ are as in (\ref{ab}). One can check that the product of such unitary matrices correctly encodes the quaternionic multiplication, and
\be
\bar{q} = q^\dagger,
\ee
where we have the quaternion conjugation on the left and the usual Hermitian conjugation on the right. The quaternionic norm is then
\be
|q|^2=\frac{1}{2} {\rm Tr}(q q^\dagger)={\rm det}(q).
\ee
Matrices of the type (\ref{quat-matrix}) satisfy $qq^\dagger = |q|^2 \id$, which is a generalised unitarity condition. We continue to refer to them as unitary. This matrix model for quaternions converts manipulations with quaternions to more familiar matrix manipulations. This can be useful. 

\subsection{The group ${\rm UH}(2)$ of unitary quaternionic matrices}

The matrix representation of quaternions also allows us to represent unitary $2\times 2$ matrices with quaternionic entries as more familiar $4\times 4$ matrices with complex entries. This way of thinking does not generalise to the octonions, because of non-associativity of the latter. But it is instructive to phrase the discussion of the previous sections in completely elementary terms of matrices, and this is why we present this viewpoint. 

Let $A,B,C,D\in \Hq$ be quaternions, which we view as $2\times 2$ matrices. Let 
\be\label{spinor-so5}
S=\left( \begin{array}{c} p \\ q \end{array} \right) \in \Hq^2
\ee
be a 2-component column with quaternionic entries $p,q\in \Hq$. We can view this as a $2\times 4$ matrix with two columns and 4 rows. The $2\times 2$ matrix with quaternionic entries
\be\label{g-so5}
g = \left( \begin{array}{cc} A & B \\ C & D \end{array}\right)
\ee
acts on 2-component rows by multiplication from the left, $S\to S^g=gS$. The group ${\rm UH}(2)$ arises as the subgroup of the quaternionic matrix group that preserves the norm
\be
|S|^2 = |p|^2+|q|^2,
\ee
i.e. as the subgroup 
\be
{\rm UH}(2)=\{ g\in {\rm GL}(2,\Hq): |S^g|^2=|S|^2 \}.
\ee
Using
\be
|S^g|^2 = \frac{1}{2}{\rm Tr}( p^\dagger A^\dagger + q^\dagger B^\dagger)(Ap+Bq) + \frac{1}{2}{\rm Tr}( p^\dagger C^\dagger + q^\dagger D^\dagger)(Cp+Dq),
\ee
and working out the consequences of the condition that the norm squared is preserved we get
\be
|A|^2+|C|^2 =1, \qquad |B|^2+|D|^2=1, \qquad B^\dagger A = - D^\dagger C.
\ee
The first two of these are real-valued, while the last equation is quaternion-valued. These equations taken together imply
\be\label{unitary-conds}
|A|^2=|D|^2, \qquad |B|^2=|C|^2, \qquad |A|^2+|B|^2=1, \qquad C = - \frac{|B|^2}{|D|^2}D B^{-1} A.
\ee
This already allows a count of the dimension of the group that arises. The objects $A,B$ are both unitary $2\times 2$ matrices, and each carries 4 parameters. They are subject to one condition $|A|^2+|B|^2=1$, which gives the number of free parameters in them as 7. The matrix $D$ is then free apart from the fact that its norm squared should be equal to the norm squared of $A$. This adds 3 more parameters. With matrices $A,B,D$ fixed the matrix $C$ is uniquely determined from the last equation. Thus, the dimension is 10, which is  the dimension of ${\rm SO}(5)$. 

\subsection{${\rm UH}(2)$ as a subgroup of ${\rm SL}(2,\Hq)$}

When we view the matrix entries of ${\rm UH}(2)$ as $2\times 2$ unitary matrices a group element of ${\rm UH}(2)$ is a complex $4\times 4$ matrix. Let us see that this matrix is in fact, of unit determinant. It is thus a subgroup of ${\rm SL}(2,\Hq)$, the group of unit determinant $2\times 2$ quaternionic matrices, which is also the conformal group of the Euclidean 4-dimensional space. The later is isomorphic (modulo $\Z_2$) to ${\rm SO}(1,5)$, and it is thus not surprising that ${\rm SO}(5)$ must sit inside. 

The determinant of a $2\times 2$ quaternionic matrix (\ref{g-so5}) is defined by viewing it as a $2\times 2$ matrix with $2\times 2$ block entries. With this interpretation we have
\be\label{det-44}
{\rm det}(g) = {\rm det}(A) {\rm det}(D-CA^{-1}B).
\ee
For $g$ with $A,B,C,D$ satisfying (\ref{unitary-conds}) we have 
\be
{\rm det}(g) = \frac{|A|^2}{|D|^2} ( |D|^2 +|B|^2)^2 = 1,
\ee
and so the matrix is unimodular. So, ${\rm UH}(2)\subset {\rm SL}(2,\Hq)$.

\subsection{${\rm SO}(5)$ as ${\rm UH}(2)$}

We can also describe an explicit homomorphism from ${\rm UH}(2)$ to the special orthogonal group in five dimensions. This uses the already familiar quaternionic model for the Clifford algebra in $\R^5$. Thus, let us view a vector in $\R^5$ as a pair $(r,{\bf x})\in \R\times \Hq$. We associate with this vector the following ${\rm End}(\Hq\oplus\Hq)$ matrix
\be\label{cliff-alg}
X(r,{\bf x})=\left( \begin{array}{cc} r \id & {\bf x} \\ {\bf x}^\dagger & - r \id \end{array}\right), \qquad {\bf x} =  \left( \begin{array}{cc} x_0 + \im x_1 & x_2+\im x_3 \\ - x_2+\im x_3 & x_0 - \im x_1  \end{array}\right), \qquad (x_0,x_1,x_2,x_3)\in\R^4.
\ee
It is easy to see that
\be
X(r,{\bf x}) X(r,{\bf x}) = (r^2 + |{\bf x}|^2) \id,
\ee
which means that the matrices $X(r,{\bf x})$ generate the Clifford algebra ${\rm Cl}_5$. Note that (\ref{cliff-alg}) is the direction generalisation of (\ref{matrix-X}) with $\C$ being replaced by $\Hq$. 

Moreover, using the interpretation (\ref{det-44}) of the determinant of $X(r,{\bf x})$ as that of a $4\times 4$ matrix we have
\be
{\rm det}(X(r,{\bf x})) = ( r^2 + |{\bf x}|^2)^2.
\ee
This means that we can obtain ${\rm O}(5)$ as the group of the transformations acting on matrices of the form $X(r,{\bf x})$ and preserving the determinant. The group ${\rm SL}(2,\Hq)$ naturally acts on $X(r,{\bf x})$ via
\be\label{SL-action}
X(r,{\bf x}) \to g X(r,{\bf x}) g^\dagger,
\ee
where $g\in {\rm SL}(2,\Hq)$ is of the form (\ref{g-so5}) with $A,B,C,D$ being quaternions. This action preserves the determinant, and preserves the property that the diagonal elements are multiples of the identity matrices. However, it does not in general preserve the property that the "trace" is zero. The subgroup of ${\rm SL}(2,\Hq)$ that preserves the zero trace condition is precisely ${\rm UH}(2)$, as is not hard to check. The homomorphism described has a non-trivial kernel, whose non-trivial element is minus the identity. Thus, we have
\be
{\rm SO}(5) = {\rm UH}(2)/\Z_2.
\ee

The story we just described is of course the direct analog of the story over $\C$, where $2\times 2$ matrices of the type (\ref{cliff-alg}) with ${\bf x}$ replaced by a complex number generate the Clifford algebra ${\rm Cl}_3$. The determinant is (minus) the $\R^3$ norm, and the group of transformations that preserves the norm is ${\rm SL}(2,\C)$. The group that preserves the tracefree conditions is ${\rm SU}(2)$. So, the quaternionic case works in precise analogy with the complex case. 

The complex version of the story has the well-known extension when the trace-free condition on the matrix $X$ is dropped. In this case one generates the Clifford algebra ${\rm Cl}_{1,3}$, and ${\rm SL}(2,\C)$ is the double cover of the Lorentz group. Similarly, we can drop the trace-free condition on matrices (\ref{cliff-alg}) and allow two arbitrary real numbers on the diagonal. This generates the Clifford algebra ${\rm Cl}_{1,5}$, the determinant is then related to the norm in $\R^{1,5}$, and the group of transformations that preserves the determinant is ${\rm SL}(2,\Hq)$, which is the double cover of the Lorentz group ${\rm SO}(1,5)$, which is also the conformal group in 4D.

\subsection{Spinor representation of ${\rm SO}(5)$}

The described representation ${\rm UH}(2)\sim {\rm SO}(5)$ in terms of $4\times 4$ matrices is essentially the spinor representation of ${\rm SO}(5)={\rm UH}(2)$. Indeed,  if we view each quaternion in $S$ given by (\ref{spinor-so5}) as a $2\times 2$ unitary matrix
\be
p = \left( \begin{array}{cc} z & u \\ -u^* & z^*\end{array}\right), \qquad q=\left( \begin{array}{cc}w & v \\ -v^* & w^* \end{array}\right), \qquad z,u,w,v\in \C,
\ee
we can write $S$ as
\be\label{S-matrix}
S = \left( \begin{array}{cc} z & u \\ -u^* & z^* \\ w & v \\ -v^* & w^* \end{array}\right).
\ee
The matrix $g$ given by (\ref{g-so5}) viewed as a $4\times 4$ matrix then acts on the $2\times 4$ matrix $S$ by multiplication from the left, and preserving this form of $S$. This action restricted to the first column is the spinor representation of ${\rm SO}(5)$ acting on its $\C^4$-valued spinors.

\subsection{The subgroup of ${\rm SO}(5)$ preserving $\Hq=\C+\C$}

We now want to determine the subgroup of ${\rm SO}(5)$ that preserves a splitting of the space of quaternions into two copies of the complex plane. A more invariant way to ask this question is to phrase it in terms of a complex structure chosen to provide the split $\Hq=\C\oplus\C$. The subgroup of $2\times 2$ quaternionic matrices that acts (from the left) on 2-component quaternionic columns and commutes with the complex structure on $\Hq$ given by $L_{\bf i}$ is the group of matrices with quaternionic entries that commute with ${\bf i}$. The quaternions that commute with ${\bf i}$ are those whose ${\bf j,k}$ components are zero. In terms of their $2\times 2$ matrix representation these are the diagonal matrices. All in all, the subgroup of ${\rm UH}(2)$ that commutes with an almost complex structure chosen is the subgroup with entries $A,B,C,D$ being diagonal matrices satisfying (\ref{unitary-conds}). 

It is not hard to see that we can always parametrise such matrices as follows
\be\label{ABCD-subgroup}
A = \left( \begin{array}{cc} e^{\im\phi} a & 0 \\ 0 & e^{-\im\phi} a^*\end{array}\right), \,\, B = \left( \begin{array}{cc} e^{\im\phi} b & 0 \\ 0 & e^{-\im\phi} b^*\end{array}\right), \,\, D=\left( \begin{array}{cc} e^{\im\phi} a^* & 0 \\ 0 & e^{-\im\phi} a\end{array}\right), \,\, C=-\left( \begin{array}{cc} e^{\im\phi} b^* & 0 \\ 0 & e^{-\im\phi} b\end{array}\right)
\ee
with $ \phi\in[0,2\pi], a,b\in C$, and $|a|^2+|b|^2=1$.
Then all the conditions (\ref{unitary-conds}) are satisfied and the corresponding matrices are in ${\rm UH}(2)$. The corresponding action on $z,w,u,v$ is as follows
 \be\label{action-su21}
 \left( \begin{array}{c} z \\ w \end{array} \right)\to e^{\im\phi} \left( \begin{array}{cc} a & b \\ - b^* & a^* \end{array}\right)\left( \begin{array}{c} z \\ w \end{array} \right), \qquad
  \left( \begin{array}{c} u \\ v \end{array} \right)\to e^{\im\phi} \left( \begin{array}{cc} a & b \\ - b^* & a^* \end{array}\right)\left( \begin{array}{c} u \\ v \end{array} \right).
  \ee
 Thus, we see that the group that preserves the split $\Hq=\C\oplus\C$ is 
 \be
 G={\rm SU}(2)\times{\rm U}(1)/\Z_2.
 \ee 
 Indeed, it is clear that the element
 \be
 (-1 , -\id ) \in {\rm U}(1)\times{\rm SU}(2)
 \ee
 corresponds to the identity element in ${\rm UH}(2)$ and thus the stabiliser subgroup is the factor group by $\Z_2$. We thus find that the spinor representation of ${\rm SO}(5)$ splits as two copies of the fundamental representation of ${\rm SU}(2)\times{\rm U}(1)/\Z_2$, the subgroup that preserves the splitting of quaternions into two copies of the complex plane. 
 
It can be checked by an explicit computation that the described ${\rm U}(1)\times{\rm SU}(2)$ subgroup of ${\rm UH}(2)$, via the homomorphism to ${\rm SO}(5)$ gets mapped into the "diagonal" subgroup ${\rm SO}(2)\times{\rm SO}(3)$. The first of these is the rotation in the $x_2, x_3$ plane, see (\ref{cliff-alg}) for notations. The group ${\rm SO}(3)$ then describes rotations in the $r,x_0, x_3$ space.

\section*{Acknowledgements} I am grateful to Michel Dubois-Violette, Ivan Todorov and John Baez for their comments on the versions of this paper. This work was partially supported by STFC grant ST/P000703/1.

\end{document}